\documentclass[pra,twocolumn,showpacs]{revtex4}

\usepackage{amssymb}
\usepackage{graphicx}
\usepackage{dcolumn}
\usepackage{bm}
\usepackage{amsmath}
\usepackage{amsfonts}

\newcommand{\be}{\begin{equation}}
\newcommand{\ee}{\end{equation}}
\newcommand{\ba}{\begin{eqnarray}}
\newcommand{\ea}{\end{eqnarray}}

\newenvironment{proof}[1][Proof]{\noindent\textbf{#1.} }{\ 
\rule{0.5em}{0.5em}}

\newtheorem{myproposition}{Proposition}
\newtheorem{mylemma}{Lemma}

\newtheorem{mycorollary}{Corollary}

\begin{document}

\title{Robustness of multi-qubit entanglement in the independent 
decoherence model}
\author{Somshubhro Bandyopadhyay}
\altaffiliation{Institute for Quantum Information Science, University of Calgary, Calgary, AB T2G 5C1, Canada}
\email{{som@qis.ucalgary.ca}}
\author{Daniel A. Lidar}
\altaffiliation{Departments of Chemistry and Electrical Engineering,
 University of Southern California, Los Angeles, CA 90089}
\email{{lidar@usc.edu}}
\affiliation{Chemical Physics Theory Group, Chemistry Department, and 
Center for Quantum Information and Quantum Control, University of Toronto, Toronto, ON M5S 3H6,Canada}

\begin{abstract}
We study the robustness of the GHZ (or ``cat'') class of multi-partite 
states under decoherence.
The noise model is described by a general completely positive map for
qubits independently coupled to the environment. In
particular, the robustness of $N$-party entanglement is studied in the 
large $N$
limit when (a) the number of spatially separated subsystems is fixed
but the size of each subsystem becomes large (b) the size of the
subsystems is fixed while their number becomes arbitrarily
large. We obtain conditions for entanglement in these two
cases. Among our other results, we show that the parity of an
entangled state (i.e., whether it contains an
even or odd number of qubits) can lead to qualitatively different
robustness of entanglement
under certain conditions. 
\end{abstract}
\pacs{03.65.Yz, 03.65.Ud, 03.67.Mn}

\maketitle

\section{Introduction}
Entanglement \cite{entanglement} is believed to be a crucial resource 
for
quantum information processing, e.g., quantum computing and
quantum communication \cite{qipandqc}. It is expected that
practical realizations of quantum protocols will involve independent or
collective manipulation of large-scale entanglement, i.e., entanglement
distributed among $N$ particles, where $N$ can be arbitrarily large.
Unfortunately, in the absence of active intervention, entanglement is
notoriously susceptible to decoherence \cite{qipandqc}. Intervention in
the form of distillation protocols \cite{distillation1, distillation2} 
and
error correcting codes \cite{qec} is capable of restoring robustness to
entanglement in the presence of decoherence. However, the efficiency of 
such
methods strongly depends on the \textit{a priori} robustness of the
entangled state in question. This raises the natural question of the 
\emph{%
inherent} robustness of entanglement. Recent work has shown that for 
certain
types of entangled states \cite{kempesimon,durbriegel,Stockton:03} or 
noisy
preparation procedures \cite{BandLidar:04} there is indeed an 
(unexpected)
inherent robustness. Here we continue this line of investigation and
consider the inherent robustness of multi-qubit entangled states under 
a
rather general model of uncorrelated decoherence.

Let us here note that the notion of robustness of entangled states has 
been used in a different context before \cite{Vidal99, 
Steiner03,SJA-MAJ} . However, in this paper by robustness we simply mean the ability of 
an entangled state to remain entangled in presence of decoherence.

Besides the obvious practical importance of studying the effect of 
decoherence
on multi-qubit entangled states, there is a fundamental interest as 
well.
Entanglement being a microscopic property, one may ask how often 
macroscopic
entanglement is realized in the physical world? In other words, when 
the
number of particles sharing an entangled state becomes very large,
entanglement truly becomes a macroscopic property of the system itself. 
At
the same time entanglement could become exponentially fragile, in the 
sense
that an arbitrary small amount of noise can destroy the complete 
coherence
between the superposed states. Yet, it is known that the set of 
separable
states is much smaller than the set of inseparable states 
\cite{Zyczkowski:98}%
, which suggests that entanglement should be relatively common. These
conflicting intuitions suggest that the question of the robustness of
entanglement is a subtle one.

The structure of $N$-party entanglement is considerably more complex 
than a
simple bipartite scenario where entanglement is only distributed among 
two
subsystems. For recent results on multipartite entanglement and it's 
measures one can see \cite{Bennett00,Blanchard01,Meyer02}. An important 
notion is the partitioning of the system into $%
2\leq M\leq N$ parties ($M$-\emph{partitioning}), where each of the $M$
parties of several particles is considered to be a single system with a
higher dimensional Hilbert space. A related notion is $M$-\emph{%
distillability}: some $M$-\emph{party pure entanglement} can be 
obtained by
local operations and classical communications (LOCC). A necessary 
condition
for $M$-\emph{distillability} is that the $M$-partitioned state is 
non-positive
under partial transposition (NPPT) \emph{across all bipartite cuts}
\cite{duretal}. Thus to obtain information about distillability of an
$N$-party 
state it suffices to study entanglement properties across all possible
bipartite cuts. It is important to realize that if there is distillable
entanglement between every pair then the whole state is $M$\textit{%
-distillable }as any multi-partite entangled state can be prepared 
given
sufficient bipartite entanglement between every pair. Note, that the
original $N$-party state not being distillable when all $N$ parties are
separated, does not rule out the state being distillable for some $M$%
-partitioning.

To address the issue whether entanglement can also be viewed as a
macroscopic property it is necessary to study the limit of large $N$. 
If we
assume a democratic partitioning for simplicity, i.e., that $N$ parties 
are
divided into $M$ groups such that $Mk=N$ where $k$ is the number of
particles in each group, then a natural question is, out of all 
possible $M$%
-partitionings, which partition exhibits maximal robustness? Does
entanglement exist when $N\rightarrow \infty $? How does large scale
entanglement behave, for instance, when the size of each partition 
becomes
macroscopic while keeping the number of partitionings fixed? Such 
questions
have been studied where the noise model was described by a depolarizing
channel \cite{kempesimon, durbriegel}. We address these questions in 
the
context of a rather general decoherence model in this work.

The structure of the paper is as follows. In Sec.~\ref{model} we 
introduce the
decoherence model: a general completely positive map under the
assumption that each qubit is independently coupled to the environment. 
In
Sec.~\ref{summary} we preview, for convenience, the main results of 
this work. Section~\ref{density} finds the
resulting density matrix under the action of decoherence. 
Section~\ref{properties} deals with
some of the useful properties of the noisy density matrix. In 
Sec.~\ref{threshold} we
obtain the entanglement  conditions that quantify the robustness of
the output state. Section~\ref{conc} concludes 
and lists some open problems.
\section{Decoherence Model}
\label{model}
In Refs. \cite{kempesimon,durbriegel} properties of large-scale 
entanglement
in the presence of the \emph{depolarizing channel} were studied. Ref. 
\cite%
{kempesimon} studied $N$-particle GHZ (also known as \textquotedblleft
cat\textquotedblright) states and compared them to W-states 
\cite{Dur:00}
and spin-squeezed states. Ref. \cite{durbriegel} studied the rather 
general
class of graph states, which includes GHZ and cluster states. In Ref. 
\cite%
{Stockton:03} the robustness of symmetric entangled states subject to
particle loss was studied. Here we focus on $N$-particle cat states, 
but
considerably generalize the decoherence model. Results corresponding to 
the
widely used depolarizing channel or dephasing channel can be reproduced 
as
special cases of our model.

An $N$-qubit cat state is of the form $\left\vert \Psi\right\rangle 
_{N}=1/
\sqrt{2}\left( \left\vert 0\right\rangle ^{\otimes N}+\left\vert
1\right\rangle ^{\otimes N}\right) $, where $\left\vert 0\right\rangle
,\left\vert 1\right\rangle $ are the $+1,-1$ eigenstates of the Pauli 
$%
\sigma_{z}$ matrix. We assume that qubits are individually coupled to 
the
environment. The action of the noisy channel on the $m$th qubit is 
described
by a completely positive map of the form:
\begin{equation}
\rho\longrightarrow\rho^{^{\prime}}=S_{m}\left( \rho\right)
=\pi_{0}\rho+\sum_{i=1}^{3}\pi_{i}\sigma_{i}^{m}\rho\sigma_{i}^{m}
\label{noisemodel}
\end{equation}
where the $\pi (t)$ are probabilities ($\pi_{i}\geq0$$,%
~\sum_{i=0}^{3}\pi_{i}=1$), and\ the $\sigma_{i}$ are the Pauli 
matrices ($%
\sigma_{1}=\sigma_{x}$, etc.). The depolarizing channel and dephasing
channel are the special cases $\pi_{i}=(1-\pi_{0})/3~$($i=1,2,3$), and 
$%
\pi_{1}=\pi_{2}=0$, respectively. By studying this rather general model 
we
hope to provide more physical insight into the factors affecting the
entanglement of a multi-qubit system in presence of decoherence.
\section{Summary of Main Results}
\label{summary}
Before launching into the analysis, let us give a brief preview of the 
results obtained. We quantify the entanglement robustness of 
$N$-particle
cat states by establishing sufficient conditions for the state to be 
entangled. Specifically, we obtain the conditions for $M$-distillability, 
for all $2\leq
M\leq N$. These conditions are obtained for both finite and infinite 
$N$.
\begin{itemize}
\item The robustness of the entangled state depends on sums and
  differences of the probabilities $\pi$ and not on the actual 
probabilities.
\item The parity of the entangled state, i.e., whether the state is 
composed
of an even or odd number of qubits, can lead to different qualitative
behavior under certain conditions.
\item Macroscopic entanglement is found to be more robust when 
distributed among higher dimensional subsystems, while keeping the number of 
spatially separated parties unchanged.
\item The most robust partition is found to be the bipartite one where 
we have equal ($N$ is even) or approximately equal ($N$ is odd) number of
qubits on both sides.
\end{itemize}
\section{Expression for Final State}
\label{density}
We begin by studying the action of the noisy channel (\ref{noisemodel}) 
on
the input state $\left\vert \Psi\right\rangle _{N}$ :
\begin{equation}
\rho_{N}\mapsto\rho_{N}^{^{\prime}}=S_{1}S_{2}...S_{N}\left( \rho
_{N}\right)
\end{equation}
The input pure state $\rho_{N}=\left\vert \Psi\right\rangle 
_{N}\langle\Psi|$
can be written as
\begin{align}
\rho _{N}& =\frac{1}{2}[(\left\vert 0\right\rangle \left\langle 
0\right\vert
)^{\otimes N}+(\left\vert 1\right\rangle \left\langle 1\right\vert
)^{\otimes N}  \notag \\
& +(\left\vert 0\right\rangle \left\langle 1\right\vert )^{\otimes
N}+(\left\vert 1\right\rangle \left\langle 0\right\vert )^{\otimes N}] 
\notag \\
& \equiv \frac{1}{2}\left[ \eta _{00}^{\otimes N}+\eta _{11}^{\otimes
N}+\eta _{01}^{\otimes N}+\eta _{10}^{\otimes N}\right] 
\end{align}

where $\left\vert i\right\rangle \left\langle j\right\vert =\eta
_{ij},i,j=0,1$.

The action of the noise channel (\ref{noisemodel}) on the operators 
making
up the density matrix is

\begin{gather}
S(\eta _{00})=a\eta _{00}+b\eta _{11}=\Lambda _{00},  \notag \\
S(\eta _{11})=b\eta _{00}+a\eta _{11}=\Lambda _{11},  \notag \\
S(\eta _{01})=c\eta _{01}+d\eta _{10}=\Lambda _{01},  \notag \\
S(\eta _{10})=c\eta _{10}+d\eta _{01}=\Lambda _{10},
\end{gather}%
where 
\begin{equation}
a=\pi _{0}+\pi _{3},b=\pi _{1}+\pi _{2},c=\pi _{0}-\pi _{3},d=\pi 
_{1}-\pi
_{2}  \label{eq:abcd}
\end{equation}%
The coefficients $a,b$ are positive whereas the coefficients $c,d$ \ 
can be
either positive or negative. Note that the grouping of $\pi_0$ with
$\pi_3$ and of $\pi_1$ with
$\pi_2$ is a result of our choice to work in the $|0\rangle ,
|1\rangle$ basis. The final density matrix can now be written as
\begin{equation}
\rho _{N}^{^{\prime }}=\frac{1}{2}\left[ \Lambda _{00}^{\otimes 
N}+\Lambda
_{11}^{\otimes N}+\Lambda _{01}^{\otimes N}+\Lambda _{10}^{\otimes 
N}\right] 
\end{equation}%
Expanding the operator terms one obtains,%
\begin{align}
\rho _{N}^{^{\prime }}& =\frac{1}{2}[\left( a^{N}+b^{N}\right) \left( 
\sigma
_{00}^{\otimes N}+\sigma _{11}^{\otimes N}\right)   \notag \\
& +\sum_{j=1}^{N-1}\left( a^{j}b^{N-j}+b^{j}a^{N-j}\right)   \notag \\
& (\sum_{p}\left\vert m_{1}^{j}m_{2}^{j}...m_{N}^{j}\right\rangle
_{pp}\left\langle m_{1}^{j}m_{2}^{j}...m_{N}^{j}\right\vert )  \notag 
\\
& +\left( c^{N}+d^{N}\right) \left( \sigma _{01}{}^{\otimes N}+\sigma
_{10}{}^{\otimes N}\right)   \label{finaldenmat} \\
& + {\rm permutations}]  \notag
\end{align}
where $\left\vert m_{1}^{j}m_{2}^{j}...m_{N}^{j}\right\rangle _{p}$ 
denotes
a ket containing exactly $j$ $0$s and the suffix $p$ denotes 
permutation.
For a given $j$ there are $\binom{N}{j}$ permutations.

Let us now define the $N$-qubit \textquotedblleft
cat-basis\textquotedblright , consisting of $2^{N}$ states. First, 
\begin{equation*}
\left\vert \Psi _{0}^{\pm }\right\rangle =\frac{1}{\sqrt{2}}\left(
\left\vert 00...0\right\rangle \pm \left\vert 11...1\right\rangle 
\right) .
\end{equation*}%
Other than these two states, we will call any other state a member of 
the%
\textit{\ }$k$-group if it is of the form:%
\begin{equation}
\left\vert \Psi _{k}^{\pm }\right\rangle =\frac{1}{\sqrt{2}}\left(
\left\vert k:0;(N-k):1\right\rangle \pm \left\vert 
k:\overline{0};(N-k):%
\overline{1}\right\rangle \right)
\end{equation}%
This notation means that the first term of the superposition state 
consists
of $k$ zeroes and $(N-k)$ ones, and the second term is obtained by 
replacing
every zero and one of the first term with one and zero, respectively. 
The
index $k\in \{1,N/2\}$ ($N$=even) or $k\in \{1,(N-1)/2\}~$($N$=odd). If 
$N$
is even, each $k$ group has $2\binom{N}{k}$ members, except when 
$k=N/2$, for
which there are $\binom{N}{N/2}$ members. If $N$ is odd, there are 
always 2$%
\binom{N}{k}$ members in every $k$-group.

The final density matrix $\rho _{N}^{^{\prime }}$ is diagonal in the 
cat
basis. This can be easily seen from the action of the noisy channel on 
the
qubits. For example, a maximally entangled state of two qubits after 
the
action of this channel becomes a mixed state which is diagonal in the 
Bell
basis. The final density matrix can be written in the following form:

\begin{align}
\rho _{N}^{^{\prime }}& =\sum_{\mu =\pm }\alpha _{0}^{\mu }\left\vert 
\Psi
_{0}^{\mu }\right\rangle \left\langle \Psi _{0}^{\mu }\right\vert  
\notag \\
& +\sum_{k,j}\left( \alpha _{kj}^{+}\left\vert \Psi 
_{kj}^{+}\right\rangle
\left\langle \Psi _{kj}^{+}\right\vert +\alpha _{kj}^{-}\left\vert \Psi
_{kj}^{-}\right\rangle \left\langle \Psi _{kj}^{-}\right\vert \right)
\label{finalstateincatdiagform}
\end{align}%
where $\alpha =f\left( a,b,c,d\right) $. The index $k$ stands for the 
group
and the index $j$ corresponds to the different states due to 
permutation of
indices that belong to the same group. Comparing the two 
representations
Eq.~(\ref{finaldenmat}) and Eq.~(\ref{finalstateincatdiagform}) of the 
final
density matrix, and noting the fact that $\left\langle \Psi
_{kj}^{+}\right\vert \rho \left\vert \Psi _{kj}^{+}\right\rangle
+\left\langle \Psi _{kj}^{-}\right\vert \rho \left\vert \Psi
_{kj}^{-}\right\rangle =\alpha _{kj}^{+}+\alpha _{kj}^{-}$ remain 
unchanged
under local depolarizing, one can show that $\alpha _{kj}^{+}+\alpha
_{kj}^{-}=\left( a^{k}b^{N-k}+b^{k}a^{N-k}\right)/2 $ which is 
independent
of $j$.

\section{Basis Condition for Entanglement and Some Useful Lemmas}
\label{properties}

For any arbitrary $N$\textit{-}qubit density matrix $\rho $ it has been
shown that the state $\rho $ is entangled if at least for one pair 
($k,j$),
the following condition is satisfied \cite{duretal},%
\begin{eqnarray}
\left\vert \left\langle \Psi _{0}^{+}\right\vert \rho \left\vert \Psi
_{0}^{+}\right\rangle -\left\langle \Psi _{0}^{-}\right\vert \rho 
\left\vert
\Psi _{0}^{-}\right\rangle \right\vert  &>&\left\langle \Psi
_{kj}^{+}\right\vert \rho \left\vert \Psi _{kj}^{+}\right\rangle 
\label{arbentcond} \\
&&+\left\langle \Psi _{kj}^{-}\right\vert \rho \left\vert \Psi
_{kj}^{-}\right\rangle   \notag
\end{eqnarray}
Accordingly we compute the above quantities for our state. First 
observe
that, 
\begin{eqnarray}
\alpha _{0}^{+}-\alpha _{0}^{-} &=&\left\langle \Psi 
_{0}^{+}\right\vert
\rho \left\vert \Psi _{0}^{+}\right\rangle -\left\langle \Psi
_{0}^{-}\right\vert \rho \left\vert \Psi _{0}^{-}\right\rangle   \notag 
\\
&=&\frac{1}{2}[\{\left\langle 00...0\right\vert \rho _{N}^{^{\prime
}}\left\vert 00...0\right\rangle +\left\langle 11...1\right\vert \rho
_{N}^{^{\prime }}\left\vert 11...1\right\rangle \}  \notag \\
&&+\{\left\langle 00...0\right\vert \rho _{N}^{^{\prime }}\left\vert
11...1\right\rangle +\left\langle 11...1\right\vert \rho _{N}^{^{\prime
}}\left\vert 00...0\right\rangle \}  \notag \\
&&-\{\left\langle 00...0\right\vert \rho _{N}^{^{\prime }}\left\vert
00...0\right\rangle +\left\langle 11...1\right\vert \rho _{N}^{^{\prime
}}\left\vert 11...1\right\rangle \}  \notag \\
&&+\{\left\langle 00...0\right\vert \rho _{N}^{^{\prime }}\left\vert
11...1\right\rangle +\left\langle 11...1\right\vert \rho _{N}^{^{\prime
}}\left\vert 00...0\right\rangle \}]  \notag \\
&=&2{\rm Re}\left\langle 00...0\right\vert \rho _{N}^{^{\prime 
}}\left\vert
11...1\right\rangle 
\end{eqnarray}
One can readily evaluate from the expression of the final density 
matrix the
value of $\alpha _{0}^{+}-\alpha _{0}^{-}$,

\begin{equation}
\Delta \equiv \left\vert \alpha _{0}^{+}-\alpha _{0}^{-}\right\vert
=\left\vert c^{N}+d^{N}\right\vert   \label{eq:Delta}
\end{equation}
and 
\begin{align}
2\lambda _{k,N-k}& \equiv \left\langle \Psi _{k}^{+}\right\vert \rho
_{N}^{^{\prime }}\left\vert \Psi _{k}^{+}\right\rangle +\left\langle 
\Psi
_{k}^{-}\right\vert \rho _{N}^{^{\prime }}\left\vert \Psi
_{k}^{-}\right\rangle =\alpha _{kj}^{+}+\alpha _{kj}^{-}  \notag \\
& =a^{k}b^{N-k}+b^{k}a^{N-k}
\end{align}%
where $\left\vert \Psi _{k}^{\pm }\right\rangle $ are the members of 
the $k$
group.

We can now state our basic result, which follows directly from 
Eq.~(\ref%
{arbentcond}).

\begin{myproposition}
The state $\rho_{N}^{^{\prime}}$ is entangled if there is at least one 
$k$
such that 
\begin{equation}
\Delta>2\lambda_{k,N-k}.  \label{eq:cond}
\end{equation}
\end{myproposition}

Suppose that there is a $k=k_{0}$ such that the above inequality is 
satisfied.
This means that the state $\rho _{N}^{^{\prime }}$ is NPPT across the
bipartite partition where $k_{0}$ parties are on one side and 
$(N-k_{0})$ on
the other [which we denote by $k_{0}:(N-k_{0})$]. Let us also note that
choosing a specific $j$ (i.e., a given permutation) implies choosing a
specific set of $k$ parties. As $\lambda $ is independent of $j$, the
entanglement condition does not depend on which parties we have chosen 
for a
given number of parties $k$ on one side of the bipartite cut.

We now prove a useful property of the final state:

\begin{mylemma}
\label{lemma1}If the state $\rho_{N}^{^{\prime}}$ \ is entangled for 
some
bipartite partition $k:(N-k)$, then it is entangled for all bipartite
partitions $m:(N-m)$ when $m>k$ and $1\leq k,m \leq N/2 (N even), 
(N-1)/2  (N  odd) $ .
\end{mylemma}

\begin{proof}
Let us first note that we are considering only bipartite partitions 
where $k (m)$ parties are on one side and $N-k (N-m)$ are on the other. 
This automatically puts the second constraint on the allowed values of 
$k,m$.
We need to prove that $\lambda _{k,N-k}>\lambda _{m,N-m}$, i.e., $%
a^{k}b^{N-k}+b^{k}a^{N-k}>a^{m}b^{N-m}+b^{m}a^{N-m}$, when $m>k$ and 
$1\leq k,m \leq N/2 (N even), (N-1)/2 , N = odd$. Without any loss of 
generality we can assume that $a > b$ because both sides are equal when 
a=b. Suppose, the inequality is not valid, i.e., $%
a^{k}b^{N-k}+b^{k}a^{N-k} \leq a^{m}b^{N-m}+b^{m}a^{N-m}$. Then 
factoring a
common $a^{k}b^{k}$ from both sides we have $a^{N-2k}+b^{N-2k} \leq 
(\frac{a}{b}%
)^{l}b^{N-2k}+\left( \frac{b}{a}\right) ^{l}a^{N-2k}$ where $m=k+l$. 
Let $\frac{a}{b} = x $. Then $x^{N-2k} +1 \leq x^{l}+x^{N-2k-l}$ , which 
implies $x^{N-2k} - x^{N-2k-l} \leq x^{l}-1$. This can be rewritten as 
$x^{N-2k-l}(x^{l} - 1) \leq x^{l}-1$  which is a contradiction because 
$x >1$ and $N-2k-l=N-(k+m) >0$ (since, $k+m<N$).
\end{proof}

It follows immediately from this lemma that:

\begin{mycorollary}
(i) If the final state is entangled across the $1:N-1$ cut then it is
entangled across all other bipartite cuts. (ii) The most robust 
bipartite
cut is $N/2:N/2$ ($N$ even) or $(N\pm1)/2:(N\mp1)/2$ ($N$ odd).
\end{mycorollary}

Let us now consider an $M$-partition with the partitions labeled as $%
G_{1},G_{2},..,G_{M}$, $2\leq M\leq N$. We call such a choice an $M$%
-partition configuration. Let $\left\vert G_{k}\right\vert $ be the 
number
of particles in the group $G_{k}$. Then an $M$-qubit cat state can be
distilled iff the state is NPPT for all possible bipartite cuts in the 
$M$%
-partition configuration. That includes the bipartite partitions 
$\left\vert
G_{k}\right\vert :$ $\left( N-\left\vert G_{k}\right\vert \right) $ 
$\forall
G_{k}$, as well as the bipartite partitions obtained by combining a 
subset
of the partitions $G_{m}$. Let $\left\vert G_{i}\right\vert =\min_k 
\left\{
\left\vert G_{k}\right\vert \right\}$. Then we have the following
lemma, that states under which condition one can distill an $M$-qubit 
cat state.

\begin{mylemma}
An $M$-qubit cat state can be distilled if and only if $\Delta>2\lambda
_{|G_{i}|,N-|G_{i}|}$.
\end{mylemma}

\begin{proof}
Using the same calculation as in Lemma~\ref{lemma1}, note that if 
$\Delta
>2\lambda _{|G_{i}|,N-|G_{i}|}$, then $\Delta >2\lambda 
_{|G_{k}|,N-|G_{k}|}$
$\forall G_{k}$. This guarantees that the final state is NPPT across 
all
other possible bipartite cuts in the $M$-party configuration. Now note 
that
our state can be brought to a depolarized form by local operations 
while
satisfying the condition $\Delta >2\lambda _{|G_{i}|,N-|G_{i}|}$ as 
well as
preserving the NPPT property. For such a depolarized state 
distillability is
guaranteed since the satisfaction of the above condition is both 
necessary
and sufficient for $M$-qubit distillability \cite{duretal}. Hence our 
state
is also distillable.
\end{proof}

Note that the symmetry properties of the state and the hierarchical 
property
of the $\lambda$'s guarantee that it is sufficient to compute the 
partial
transposition of only one bipartite cut. The next lemma shows the 
connection
between a bipartite cut and the largest $M$ one can have in $M$-qubit
distillability. Let $Q(N,k)=$quotient of $(N-k)/k$.
\begin{mylemma}
For any given $k$ in a bipartite cut $k:(N-k)$, if the distillability
condition $\Delta>2\lambda_{k,N-k}$ is satisfied, then the density 
matrix $%
\rho_{N}^{\prime}$ is $M$-qubit cat state distillable, where $\max 
M=N/k$ if 
$N$ is divisible by $k$ or $\max M=1+Q(N,k)$ if $N$ is not divisible by 
$k$.
\end{mylemma}
\begin{proof}
Let $k$ be the smallest integer such that $\Delta >2\lambda _{k,N-k}$.
Clearly if $N$ is divisible by $k$, the maximum number of groups one 
can
have is $N/k$. However if $N$ is not divisible by $k$, then the 
remainder of 
$N/k$ is less than $k$ and hence $1+N/k$ cannot be the number of 
optimal
groups, as the partition (\textit{remainder of }$N/k)$ :($N-$\textit{%
remainder of }$N/k)$ is not NPPT. Therefore the optimal number of 
partition
must be $1+Q(N,k)$.
\end{proof}

It is clear that every bipartite cut contains distillability 
information
about some $M$-partition. In what follows we will analyze the 
entanglement
threshold conditions for bipartite cuts. The previous lemmas guarantee 
that
the bipartite threshold conditions are sufficient to describe 
distillability
of the final state $\rho_{N}^{\prime}$.
\section{Entanglement properties of $\rho_{N}^{^{\prime}}$}
\label{threshold}
In this section we study in detail entanglement properties of the state 
$\rho_{N}^{^{\prime}}$. As noted before we will concentrate on the 
conditions for the state to remain entangled across the bipartite cuts. Let 
us first point out the effect parity (in the sense of odd/even number
of qubits) can have on the inequality. As mentioned before, the
parameters $c,d$ can be either positive or negative. Without loss of
generality suppose $\left\vert c\right\vert \geq\left\vert d\right\vert 
$.
When $c,d$ have the same sign we have [recall Eq.~(\ref{eq:Delta})] $%
\Delta=\left\vert c\right\vert ^{N}+\left\vert d\right\vert ^{N}$
irrespective of $N$ being even or odd. However if $c,d$ have opposite 
signs
then $\Delta=\left\vert c\right\vert ^{N}+\left\vert d\right\vert ^{N}$ 
when 
$N$ is even or $\Delta=$ $\left\vert c\right\vert ^{N}-\left\vert
d\right\vert ^{N}$ when $N$ is odd. This suggests that parity can have 
a
dramatic effect on the sufficient condition for entanglement. 
%More specifically,
%as we now show in detail, for a finite number of qubits this sufficient 
 %condition for an odd number of qubits can be considerably tighter
%than for states with even number of qubits. 
%Entanglement for odd-qubit
%states can even vanish, while even-qubit states may show some entanglement
%when $\left\vert c\right\vert =\left\vert d\right\vert $.

\subsection{Sufficient condition for entanglement  when $c,d$ have the 
same sign}

\subsubsection{Exact condition for finite N}

Let us rewrite our basic NPPT condition (\ref{eq:cond}) explicitly as 
$%
(\left\vert c\right\vert ^{N}+\left\vert d\right\vert
^{N})>a^{k}b^{N-k}+b^{k}a^{N-k}$. Let us note first that, since from 
Eq.~(%
\ref{eq:abcd}) it follows that $a\geq c$ and $b\geq d$, the NPPT 
condition is
violated when $a=b$.

%\begin{myproposition}
%A necessary condition for satisfying Eq. (14) is $a\neq b$.
%\end{myproposition}

From hereon we assume $a\neq b$. Note that if $\left\vert c\right\vert
=\left\vert d\right\vert $ the entanglement condition can still be 
satisfied
if $a\neq b$. Letting $k=N\alpha$, where 
$\frac{1}{N}\leq\alpha\leq\frac{1}{2%
}$, one obtains $\left\vert c\right\vert ^{N}(1+\left\vert \frac{d}{c}%
\right\vert ^{N})>b^{\alpha N}a^{(1-\alpha)N}(1+\left( 
\frac{b}{a}\right)
^{(1-2\alpha)N})$, where one can assume without loss of generality that 
$a>b$
and $\left\vert c\right\vert >\left\vert d\right\vert $. Taking the
logarithm of both sides and dividing by $N$, one obtains

\begin{align}
\log \left\vert c\right\vert +\frac{1}{N}\log (1+\left\vert 
\frac{d}{c}%
\right\vert ^{N})& >\log \left( b^{\alpha }a^{(1-\alpha )}\right)   
\notag \\
& +\frac{1}{N}\log (1+\left( \frac{b}{a}\right) ^{(1-2\alpha )N}).
\label{entcondition1}
\end{align}%
This inequality is an exact sufficient condition for entanglement 
corresponding to a bipartite cut $k:(N-k)$.

\subsubsection{Asymptotic  condition when the subsystems and
  system sizes become macroscopic}

Consider first the case when the subsystem size $k$ becomes macroscopic 
in
the limit $N\rightarrow\infty$. This implies $\alpha$ remaining a 
constant
as $N\rightarrow\infty$. The asymptotic condition can easily be 
obtained
from Eq.~(\ref{entcondition1}):

\begin{equation}
\left\vert c\right\vert >b^{\alpha}a^{(1-\alpha)}  
\label{entcondition3}
\end{equation}
Note that $d$ dropped out in the asymptotic limit. We can rewrite this 
as a
lower bound on $\alpha$ in the asymptotic limit, using the 
normalization
condition $a+b=1$:

\begin{equation}
\alpha>\frac{\log\left( \frac{a}{|c|}\right) }{\log\left( 
\frac{a}{1-a}%
\right) }\equiv f(a,|c|)  \label{entcondition2}
\end{equation}
Remarkably, the asymptotic  condition depends only on \emph{two}
parameters, $a$ and $c$. Noting that $\alpha=1/M$, 
Eq.~(\ref{entcondition2})
puts an upper bound on $M$.

Since $\frac{1}{N}\leq\alpha\leq\frac{1}{2}$ we find from Eq.~(\ref%
{entcondition3}) a particularly simple form of the entanglement 
condition in
the case of the most robust partition, i.e., the case 
$\alpha=\frac{1}{2}$:

\begin{equation}
\left\vert c\right\vert ^{2}>ab
\end{equation}

We note here that an identical inequality can also be obtained by 
putting
the constraint $f(a,|c|)\leq 1/2$. 

\subsubsection{Robustness and size of the subsystems}

An interesting feature of Eq.~(\ref{entcondition1}) is that we can 
separate
out the asymptotic part from the $N$-dependent term. Let us rewrite 
Eq.~(\ref%
{entcondition1}) as:%
\begin{align}
\log \left\vert c\right\vert & >\log \left( b^{\alpha }a^{(1-\alpha
)}\right)   \notag \\
& +\frac{1}{N}\left[ \log (1+\left( \frac{b}{a}\right) ^{(1-2\alpha
)N})-\log (1+\left\vert \frac{d}{c}\right\vert ^{N})\right] 
\label{thresholdineq2}
\end{align}%
In this form it is easy to see that the asymptotic condition is 
achieved
even for finite $N$ if and only if $\log (1+\left( \frac{b}{a}\right)
^{(1-2\alpha )N})=\log (1+\left\vert \frac{d}{c}\right\vert ^{N})$, 
i.e.,
the following condition is satisfied:

\begin{equation}
\alpha=\frac{1}{2}\left( 1-\frac{\log\left\vert \frac{c}{d}\right\vert 
}{%
\log\left( \frac{a}{b}\right) }\right) .  \label{eq:alpha-finite}
\end{equation}
For instance, when $N$ is even, for a choice of $\left\vert 
c\right\vert
=\left\vert d\right\vert $, one obtains $\alpha=1/2$. This shows, 
rather
surprisingly, that the asymptotic  condition is the same as that of
any finite $N$ if certain conditions are met. Consistency demands that, 
in
such a case, Eqs. (\ref{entcondition2}),(\ref{eq:alpha-finite}) should 
be
satisfied simultaneously. This leads to the condition $|c||d|>ab$, 
which
cannot be satisfied in general, except for the case when $\left\vert
c\right\vert =\left\vert d\right\vert $. We therefore conclude that 
latter
surprising property holds only for equi-grouped partition, and only 
when $N$
is even, because $\alpha\neq1/2$ if $N$ is odd. This therefore shows 
another
effect of parity of the number of qubits.

We now examine Eq.~(\ref{thresholdineq2}) more closely. Let us denote 
$%
\mu\equiv\log(1+\left( \frac{b}{a}\right) ^{(1-2\alpha)N})-\log
(1+\left\vert \frac{d}{c}\right\vert ^{N})$ and rewrite Eq.~(\ref%
{thresholdineq2}) as

\begin{equation}
\log \left\vert c\right\vert >\log \left( b^{\alpha }a^{(1-\alpha 
)}\right) +%
\frac{\mu }{N}.  \label{thresholdineq3}
\end{equation}%
Let us first observe that $\mu \geq 0$ if $\alpha \geq 
\frac{1}{2}\left( 1-%
\frac{\log \left\vert \frac{c}{d}\right\vert }{\log \left( \frac{a}{b}%
\right) }\right) $. Therefore if Eq.~(\ref{thresholdineq3}) is 
satisfied for
some finite $N$ when $\mu $ is positive, then the asymptotic 
condition of entanglement (\ref{entcondition2}) is automatically 
satisfied
as well. As before, demanding the consistency requirement one can show 
that $%
\frac{1}{2}\left( 1-\frac{\log \left\vert \frac{c}{d}\right\vert }{\log
\left( \frac{a}{b}\right) }\right) <f(a,|c|)$ if $|c||d|<a b$, which, 
in
general, is always satisfied. We can therefore conclude that existence 
of $M$%
-group entanglement for some finite $N$ automatically implies 
entanglement
also in the asymptotic limit when the size of the partitions become
macroscopic, keeping the number of partitions $M$ a constant. On the 
other
hand if the inequality (\ref{thresholdineq3}) is not satisfied for some 
$N$,
it might still be satisfied for some large $N$, as the condition itself 
gets
relaxed as $N\rightarrow \infty $. We summarize these considerations as
follows:

\begin{myproposition}
Multi-partite entangled states with fixed number of partitions $M$ are 
more
robust the larger is the dimension of the constituent subsystems.
\end{myproposition}

\subsubsection{Fixed number of members in a group while number of 
groups and
the size of the system become large}

We now discuss the second scenario where we allow the number of 
particles in
a group to remain constant, i.e., $k$ is fixed while both $M,N$ are 
allowed
to become arbitrarily large. For simplicity we assume that each group
contains the same number of qubits. We first rewrite 
Eq.~(\ref{thresholdineq2})
as 
\begin{align}
\log\left\vert c\right\vert & >\frac{k}{N}\log b+(1-\frac{k}{N})\log a 
\notag \\
& +\frac{1}{N}\left[ \log(1+\left( \frac{b}{a}\right)
^{(N-2k)})-\log(1+\left\vert \frac{d}{c}\right\vert ^{N})\right]
\end{align}
Suppose the above inequality is satisfied for some choice of $k,N$. 
However,
in the limit $N\rightarrow\infty$ the inequality reduces to the
condition $\log\left\vert c\right\vert >\log a$, which is false. Thus:

\begin{myproposition}
If we allow both the number of partitions and the number of qubits to 
become
large, inequality (22) ceases to be satisfied.
\end{myproposition}

\subsection{Condition for entanglement  when $c,d$ have opposite signs}

Let us now consider the case of odd number of qubits and $c,d$ having
opposite signs, whence $\Delta=\left\vert c\right\vert ^{N}-\left\vert
d\right\vert ^{N}$. First, note that the condition (\ref{eq:cond})
is considerably tighter than before. Second, there is no distillable 
entanglement in
this case when $c=d$ as opposed to the same-sign case, irrespective of 
whether 
$a,b$ are equal or not.

Let us write the  condition (\ref{eq:cond}) in this case as:

\begin{align}
\log \left\vert c\right\vert & >\log \left( b^{\alpha }a^{(1-\alpha
)}\right)   \notag \\
& +\frac{1}{N}\left[ \log (1+\left( \frac{b}{a}\right) ^{(1-2\alpha
)N})-\log (1-\left\vert \frac{d}{c}\right\vert ^{N})\right] 
\end{align}%
In the asymptotic limit, we obtain the same condition as before.
However, notice that the term in the parentheses on the right side of 
the
inequality is always positive. This implies that as we increase the 
number
of particles robustness always increases for any choice of $\alpha $. 
This
means that if the state is entangled for some choice of $M,N$, it 
always
remains entangled in the large $N$ limit, as robustness increases with 
$N$
as long as the number of partitions $M$ remain unchanged.

\section{Conclusions and open problems}
\label{conc}

To summarize, in this work we have studied the robustness of $N$-qubit 
cat
states under a rather general decoherence model. In this model every 
qubit is
independently coupled to the environment. The noisy channel is 
described by a completely positive map
with arbitrary probabilities assigned to the various errors described 
by the
Pauli operators. Our findings show that macroscopic entanglement is 
more
robust in higher dimensional systems while keeping the number of 
spatially
separated parties constant. We have also shown that states with even or 
odd
numbers of qubits can have qualitatively different properties that are 
not
observed in simpler noisy channels such as depolarizing channel. 
Furthermore,
we have shown that in the asymptotic limit the entanglement 
condition depends only on two noise parameters even though the noise 
model
itself is described by three independent parameters.

The present work focuses on the GHZ class of states. It would be
interesting to see if the above observations hold true for other 
classes of
multi-qubit pure states. Kempe and Simon \cite{kempesimon} have studied
the robustness of W-states for three qubits, and other inequivalent 
classes of
four-qubit entanglement, using a depolarizing channel. Their work has
shown that GHZ states are more robust than the other classes of
states. It would be interesting to test whether this holds true under
the more general noise model considered here. The same comment applies
to the class of graph states studied by D\"{u}r and Briegel 
\cite{durbriegel}.

Note added: Upon completion of this work we came to know about related
work by Hein {\it et al.} \cite{hein:04}, which uses a Markovian master 
equation approach to
model decoherence, whereas in this work we have used the (formally 
exact) Kraus operator sum
representation.

\section*{Acknowledgement}
Financial support from the Sloan Foundation, PREA and NSERC
(to D.A.L.) is gratefully acknowledged.

\end{document}